\documentclass[12pt,preprint]{aastex}
\usepackage{epsfig}
\usepackage{amssymb}
\usepackage{graphicx}
\makeatletter

\usepackage{natbib}
\usepackage{fancyhdr}

\makeatother

\shorttitle{First Super-Earth's structure}
\shortauthors{Valencia et al.}

\begin{document}

\title{Radius and Structure Models of the \\
first Super-Earth Planet}

\author{Diana Valencia\altaffilmark{1}}
\affil{Earth and Planetary Science Dept., Harvard University, 20 Oxford Street, Cambridge, MA, 02138}
\email{valencia@mail.geophysics.harvard.edu}

\author{Dimitar D. Sasselov}
\affil{Harvard-Smithsonian Center for Astrophysics,
60 Garden Street, Cambridge, MA, 02138}

\and
 
\author{Richard J. O'Connell}
\affil{Earth and Planetary Science Dept., Harvard University, 20 Oxford Street, Cambridge, MA, 02138}

\altaffiltext{1}{corresponding author}




\begin{abstract}
With improving methods and surveys, the young field of extrasolar
planets has recently expanded into a qualitatively new domain - terrestrial
(mostly rocky) planets. The first such planets were discovered during
the past year, judging by their measured masses of less than 10 Earth-masses
($M_{\oplus}$) or Super-Earths. They are introducing a novel physical
regime that has not been explored before as such planets do not exist
in our Solar System. Their composition can be either completely terrestrial or
harbour an extensive ocean (water and ices) above a rocky core.
We model the structure and properties of the
first Super-Earth (mass $\sim$ 7.5 $M_{\oplus}$) discovered in 2005,
illustrating the possibilities in composition and providing radius
evaluations in view of future detection of similar planets by transits. We
find that a threshold in radius exists for which larger values indicate that
a Super-Earth most certainly has an extensive water content.  In the case of
GJ876d this threshold is at about 12000 km. Our results show that unique
characterization of the bulk composition of Super-Earths will be possible in
future transit studies.

\end{abstract}


\section{Introduction}
In the last 11 years more than 200 exoplanets have been discovered but only in the last year the minimum mass for
detection has been pushed down to allow for terrestrial planets.
The first extra-solar planet with a mass lower than 10 $M_{\oplus}$,
i.e. a Super-Earth, was discovered last year by \citet{Rivera_et_al:2005} 
orbiting an M4V star named GJ876. Its estimated mass
is 7.5$\pm$0.7 $M_{\oplus}$ and it has an orbital period of 1.94
days. It is close to the host star and the surface temperature
is calculated to range between 430 and 650 K \citep{Rivera_et_al:2005} . The star is
known to have two gas giant planets GJ876b,c in 30 and 61 day orbits
\citep{Marcy_et_al:1998, Marcy_et_al:2001}. Hence, the planetary system of star GJ876 has both an
architecure and a planet - the Super-Earth, that are unfamiliar to
our Solar System experience.  Two other low mass planets have been discovered
in this year:
OGLE-2005-BLG-390Lb ($\sim 5M_{\oplus}$ at 5 AU --- \citep{beaulieu:2006}) and HD69830b ($\sim
10M_{\oplus}$ at 0.08 AU --- \citet{Lovis_et_al:2006} ) exemplifing the variety in Super-Earths
that will be discovered in the near future with missions like \emph{Kepler}.

Calculating the internal structure of
a Super-Earth can help us determine how different or similar to
the Earth this planet might be. One of our goals is to provide mass-radius
relations that will help characterize Super-Earths discovered by the transit
method. We show in this study the tools we have to
characterize Super-Earths and use GJ876d as an example of our model
capabilities and limitations.
We first describe the numerical method used to obtain density, pressure, and temperature profiles
as a function of radius and refer the reader for more details to
\citet{Valencia_et_al:2006},
 where we provided the first theoretical grid of models for Super-Earth planets. 
To calculate the planet's radius and internal structure we
need to make reasonable assumptions about its composition. It might be
completely rocky or might have
accumulated a substantial amount of ices depending on the material available
during formation. We show the results and implications
for different likely compositions for GJ876b as an example that can be
replicated for any Super-Earth.
Additionally, we explore the effects of tidal heating that might be present in short period
orbits (like GJ876d) and nonzero eccentricities.
\\
\section{Model}
\subsection{Numerical method and Equations of State}
We model the planet as being composed of distinct, homogeneous in composition, spherical
shells.  For a terrestrial planet these shells are: the mantle that is divided into 
lower mantle and upper mantle, and the core that depending on the temperature
structure of the planet might be divided into a liquid outer core and a solid
inner core, as is the case for the Earth. The composition for the different
layers is taken from the mineralogical composition known for the Earth \citep{McDonough_Sun:1995}. The
upper mantle is composed of olivine (ol) and higher pressure forms of olivine
(wadsleyite (wd) and ringwoodite (rw)); the lower mantle develops when rw
transforms to perovskite (pv) and ferromagnesiowustite (fmw), with an
additional shell at high
pressures when pv transforms to post-perovskite (ppv). The solid inner
core in the Earth is composed of Fe and small quantities of Ni, and the outer
liquid core is composed of Fe and a light alloy (see figure 1). The candidates for this
alloy are S, Si, O, C and H and to this day there is no consensus on which one(s)
or the amount(s). We have used in this study a composition of pure Fe and
Fe$_{0.8}$(FeS)$_{0.2}$ to show the uncertainties in the radius of a planet from
the lack of knowledge of the composition of the core.\\

For planets that harbour a substantial H2O/ice content (ocean planets), there is an additional layer overlying the rocky
interior (mantle and core), composed of two shells - water above high
pressure phases of ice.
In some cases pressure is not high enough (amount of H$_2$O is not large
enough) to solidify water and the ice layer is absent (figure 1).  The thickness of the water is determined by the intersection
between the PT curve of the planet and the melting curve of ice. In the case
of low surface temperature, the planet will have an additional shell above the
water layer composed of ice I --- the lightest form of H$_2$O and negative Clapeyron
slope -- as is the case for Jupiter's satellites.\\

The numerical model solves the following differential equations for density $\rho$, gravity
$g$, mass $m$ and pressure $P$ in each shell with a Runge-Kutta 4th order solver:

\begin{eqnarray}
\frac{d\rho}{dr} & = & -\frac{\rho(r)g(r)}{\phi(r)}\\
\frac{dg}{dr} & = & 4\pi G\rho(r)-\frac{2Gm(r)}{r^{3}}\\
\frac{dm}{dr} & = & 4\pi r^{2}\rho(r)\\
\frac{dP}{dr} & = & -\rho(r)g(r)\end{eqnarray}
 
where $\phi(r)=\frac{K_S(r)}{\rho(r)}$ is the seismic parameter that can be
calculated with an equation of state (EOS) for $K_S$, the adiabatic bulk
modulus; $G$ is the gravitational constant and $r$ is the radius.

The model integrates from the surface inwards with boundary conditions for
P, T, $\rho$ and total mass (M). It starts with a guess for the planet's radius (R) that yields
a surface gravity value and determines the structure to the last shell where
there could be excess or deficit in mass depending whether the starting R is too small
or too big respectively.  The model then uses a bisect-newton
approach to determine the radius that yields zero mass and zero gravity
at the center of the planet (within 5 meters).

The EOS that we have implemented for all layers except the water layer is the Vinet EOS, because is the best analytical
EOS for extrapolation \citep{Hama_Suito:1996}:

\begin{equation}
K_T (\rho,300) =3K_0\left( x^{2/3}-x^{1/3}\right) \exp{\left(\frac{3}{2} (K_0^{\prime}-1)\left(1-x^{-1/3}\right)\right)}
\end{equation} 

where $x=\frac{\rho}{\rho_0}$, $K_0$ and $K_0^{\prime}$ are the
isothermal bulk modulus and its first derivative at a reference state --- zero pressure and 300
Kelvin --- and $ K_S=K_T \left(1+\alpha \gamma T \right)$, where $\alpha$ is the
thermal expansion coefficient and the Gruneisen parameter $\gamma=\gamma_0
x^{-q}$ 
with $q=-\frac{\partial\ln\gamma}{\partial\ln\rho}$. Most high-pressure
experiments fit their data to the 3rd order Birch-Murnaghan EOS (BM3) that is
derived from a truncation of the linear expansion of the Helmholtz free energy with
strain to the third order \citep{Poirier:Earths-interior:EOS}. The extrapolation of this EOS is highly uncertain since the fourth
term in the expansion might not be smaller than the 3rd term in some cases.
We use the values for $K_0$ and $K_0^\prime$ obtained from the literature corresponding to the BM3 EOS to
obtain P(V) and refit the data to the Vinet EOS within a volume compression
between 1 and 2/3. The general trend is that $K_0$ is smaller by a few GPa and
$K_0^\prime$ is larger by $\sim$6-8\% in the Vinet fit. We used the
Rangkine-Hugoniot EOS from \citet{Stewart_Ahrens:2005} within the water region.\\

To incorporate the effects of temperature in the EOS we
added a thermal pressure term (due solely to deviations in temperature
from 300K) for core and mantle regions that translates into a thermal bulk
modulus correction.  With $R$ the universal gas constant, $\theta=\theta_0
\exp\left({\frac{\gamma_0-\gamma}{q}}\right)$ as the Debye temperature and $n$ as the
number of atoms in the unit cell, the thermal correction is expressed as:
\begin{eqnarray}
 K_T(\rho,T) &  = & K_T(\rho,300)+\Delta K_{th}(\rho,T)\\
\Delta K_{th} & = & 3nR\gamma\rho\left(f(T)-f(T_{0})\right)
\end{eqnarray}
\begin{equation}
f(T) =\left(1-q-3\gamma\right)\frac{T^{4}}{\theta^{3}}\int_{0}^{\frac{\theta}{T}}\frac{\xi^{3}d\xi}{\exp\xi-1}+3\theta\gamma\frac{1}{\exp\left(\theta/T\right)-1}
\end{equation}

For the icy region we incorporated the temperature effects in the density via
the thermal expansion coefficient:
\begin{eqnarray}
\rho(P,T) & = & \rho(P,300)\exp\left[\int_{300}^T \alpha(P,T')dT'\right]\\
\alpha(P,T) & = & (a_0+a_1T)\left( 1+\frac{K_0^\prime}{K_0}P\right)^{-b}
\end{eqnarray}
where $a_0$, $a_1$ and $b$ are coefficients determined at zero pressure.

\subsection{Thermal model}
In order to incorporate the temperature effect in the EOS we need to have a model
describing the temperature profile in the planet. We use the Earth as our
starting point and parameterized convection theory to model the temperature
regime of Super-Earths. The surface heat flow on Earth (44TW \citet{Pollack:1993})reflects the heat from 
radioactive sources in the mantle (some authors also place
potassium in the core) and secular cooling
from an initial hot state. If we account for the concentration of heat sources
in the crust (generating about 5-12TW
\citep{Davies:Dynamic-Earth,Vacquier-1991,Taylor-McClennan:1998}, 
we take 8TW), and assume a bulk silicate concentration
of uranium, potassium and thorium, radioactive heat in the mantle accounts for $\sim
58\%$ of the total heat flow. In \citet{Valencia_et_al:2006} we reported scaling the
heat flow with mass for Earth-like planets as a first attempt to
scale radioactive heat sources and secular cooling together.  Here we refine this
scaling by separating the treatment of the two sources.
If we assume that the material that makes Super-Earths is the
same as the bulk material for the Earth, then the radioactive heat sources
scale with mantle mass. Planets that have larger mantles will have higher heat
flows.  It is difficult to scale secular
cooling for a Super-Earth because we do not know the role that mass plays in the
evolution of a planet.  With the scaling laws derived in
\citet{Valencia_et_al:2006} and parameterized convection, we have found
(Valencia et al, \emph{in prep}) that a massive planet is likely to convect in a plate
tectonic regime similar to the
Earth. Intuitively, the more massive the planet is, the higher the Rayleigh
number that controls convection, the thinner the top boundary layer
(lithosphere) and faster the convective velocities.   This scenario might aid 
subduction of the lithosphere causing the onset of plate tectonics.  Thus, we
adopt that the evolution of a Super-Earth might be one that leads to the same
proportion of secular cooling to radioactive heating as in the Earth.  This is the first
attempt to scale secular cooling with mass. \\

We define the Rayleigh
number $Ra$ (dimensionless parameter that controls convection) in terms of the
heat flux ($q_s$):

\begin{equation}
Ra=\frac{\rho g\alpha q_{s}/k}{\kappa\eta}D^{4}\label{eq:Ra}
\end{equation}

where $g$, $k$, $\kappa$, $\eta$ are the average gravity, conductivity,
diffusivity and viscosity of the mantle. The thickness of the top boundary layer is   
\begin{equation}
\delta=a\frac{D}{2}\left(\frac{Ra}{Ra_{crit}}\right)^{-1/4}\label{delta}
\end{equation}
independent of the size of the mantle, where $Ra_{crit}\sim 1000$ is the critical Rayleigh
number and $a$ is a coefficent of order unity. We treat viscosity in two ways: i) isoviscous case
and ii) a temperature dependent treatment $\eta(T)=\eta_0 (\frac {T}{T_0})^{-30}$
(see \citet{Valencia_et_al:2006} for more details).\\

The planet modeled in this study has a convective core, mantle and ice/water
layer where the temperature can be described as adiabatic:
\begin{equation}
\frac{dT(r)}{dr}=\frac{\rho(r)g(r)T(r)}{K_S(r)}\gamma(r)\label{eq:adiabatic gradient}\end{equation}

All interfaces between the chemically distinct layers develop boundary layers that transfer heat conductively 
\begin{equation}
\frac{dT}{dr}=-\frac{q}{k}\label{eq:fouriers law}
\end{equation}

Their thickness depends strongly on the local viscosity. Owing to the low
viscosity of the core and water we only model conductive boundary layers at
the top and bottom of the mantle.  The top boundary layer thickness is
described in equation (12). In an isoviscous internally heated case, the bottom boundary
layer of the mantle would vanish. Conversely in an isoviscous heated-from
below case it
 would have the same thickness as the top one.  Since we
assume these planets are in a intermediate regime (same as Earth) the bottom
boundary layer is taken to be half of the top one, and the heat flux reflects
the heat coming out of the core.

Even though there are two critical assumptions in determining the temperature
structure --- the constant ratio between internal heating and secular cooling,
and the thickness of the lower boundary layer in the mantle --- the effects of
temperature in the internal structure of a planet are small, particularily the
effects in radius. The temperature profile is needed mostly to determine the location
of the phase changes that determine the different regions in the planet. A
temperature dependent viscosity increases the radius by $\sim 20$ km compared
to the iso-viscous case showing that neither assumption is critical.

The numerical model iterates several times until
convergence exists between the boundary layer thickness and the radius (i.e.
surface heat flux) so that the temperature profile is determined
self-consistently.

The numerical model needs as input the composition of the different regions as
well as the proportion of ice mass fraction (IMF) and core mass fraction
(CMF), and surface P and T.  The output is: density,
pressure, temperature, mass and gravity as a function of radius; the total
radius and the location of the phase changes and composition boundaries.
Table 1a shows the different values in composition used in this study. The composition assumes
that the mantle minerals have 10\% of iron and 90\% magnesium in the mineral
structures, except for post-perovskite phase that is believed to incorporate more
Fe \citep{Mao:2004} where we adopt a 20\% Fe by mol composition.  Table 2 shows the
values we used to determine the phase boundaries between the major silicate
phase changes (ol to wd, rd to pv+fmw, pv to ppv), the phase boundary between water
and ice is determined from a Krauss-Kennedy melting curve proposed by
\citet{Stewart_Ahrens:2005}.  We calculate the melting point of iron by using
a Lindeman equation (see \citet{Valencia_et_al:2006} for details) only to test
whether the core is in a solid or liquid phase but not to distinguish between
different phases within the core (as is the case for the Earth). Only the
Earth-like composition yields a planet hot enough to have a liquid core (see
figure 3). A
planet with more water content (IMF=40\%) will have a smaller mantle for a
given CMF and will be cooler (due to less radioactive heating) than a less water rich planet (IMF=20\%).  The
strong dependency of the Gruneisen parameter with volume in the mantle (pv or ppv) (that determines the adiabatic
gradient) prevents the temperature at the core-mantle boundary of being hot
enough in planets with medium to small mantles (ocean planets with medium
cores and rocky planets with large cores). The adiabatic gradient within the
mantle remains small and the temperature increase within the mantle is not
sufficient to allow for a liquid core (figure 3).

\subsection{Composition cases for GJ876d}
We consider a few basic scenarios for the composition of GJ876d that are representative
but probably not exhaustive. They are (1) a simple Earth-like composition, a planet with a core
that makes up 33\% of the total mass, with a mantle that is 10\% (by
mol) iron enriched, and a lower mantle composed of 30\% fmw
and the rest pv.
(2) an Earth-like, but with a very large core (mass fraction of 80\%),
(3) an ocean planet with a 20\% by mass water/ice layer on top of
a terrestrial (Earth-like) core, and (4) an ocean planet with a 40\%
water/ice layer. Our reasons for introducing the later three scenarios are:
case (2) is analogous to Mercury - with its very close orbit, GJ876d
might have similarly acquired a massive core during its formation
- this case is essentially a Super-Mercury (see \citet{Valencia_et_al:2006} for details).
Cases (3) and (4) are inspired by the possibility \citep{Zhou_et_al:2005} that
GJ876d actually formed beyond the snow line in the protoplanetary
disk of GJ876.  The snow line in a disk is the distance from the star
at which the local temperature and pressure in the midplane allow
ices to exist during planet formation \citep{Sasselov_snowline:2000}. In our Solar System
the snow line was at about 2-3 AU, but GJ876 is a small 0.32 $M\odot$ 
star \citep{Rivera_et_al:2005}. With its lower energy output and smaller
size, GJ876 would define a snow line at $\sim$0.25 AU in its passive
disk at zero age. Beyond the snow line, ices (mostly H$_{2}$O) increase
the solid fraction in the disk by a factor of $\sim4$, given the
solar composition of the star. Therefore a Super-Earth forming in
this region of the disk could acquire and retain water as a major fraction
(20\% or more) of its total mass. Differentiation will
quickly produce an Earth-like core and interior overlaid by water
ice (higher pressure ice phases such as VII and X) and a deep water ocean, similar
to the ocean planets described by \citet{Leger_et_al:2004}. A more
complicated issue is whether this water could evaporate during the
lifetime ($\sim$6 Gyr) of the planet. A simple thermal escape estimate
precludes that. It is possible to retain an icy/liquid layer over
the age of this planetary system with a surface temperature of $\sim$550
K provided the atmospheric pressure is large enough. Ice VII is stable
at such temperature given pressures larger than a few GPa \citep{Stewart_Ahrens:2005}
. Furthermore, the atmospheric partial pressure would have to exceed
$\sim$100 MPa to prevent the water layer from evaporating \citep{Wagner_Pruss:water_2002}.
A non-thermal escape calculation is beyond the scope
of this paper; it could prove very efficient in removing a lot of
water from GJ876d, especially during the early evolution of the star.
\\

Addressing the issue of bulk planet composition is central to our
study. The case of GJ876 provides us with useful constraints.
First, we know \citep{Rivera_et_al:2005} that the star has a composition
very similar to the Sun and is of similar age. Second, there is evidence
that the orbits of the GJ876 planets evolved shortly after planet
formation by inward migration - the
very close orbits of all 3 planets and the 1:2 mean-motion resonance
state of the giant planets \citep{Zhou_et_al:2005} \\

\section{Results}
For the Earth-like composition case (lines with stars in top figure 2) perovskite transforms to
post-perovskite at a radius of $\sim 9800$ km, so most of its mantle is actually
composed of ppv+fmw. This brings attention to this relatively newly
discovered phase present in the lower-most mantle on Earth (only between 125
to 136 GPa) and the importance of determining with accuracy its behaviour
at high pressures. If this planet had a pure Fe core, its radius would extend to 10800 km or 1.70 $R_{\oplus}$
(Earth radius) and an additional 130 km (1.71$R_{\oplus}$) if the core was composed of
Fe$_{0.8}$(FeS)$_{0.2}$. Despite the large differences in core density (300
kg/m$^3$) between the two core compositions, due to the
different locations of the core-mantle boundary that satisfy the CMF, the
radius differs by very little. This implies that the radius for planets with medium sized
cores is a robust parameter.\\
 
The second scenario looks at the effect of having
a larger CMF of 80\% (see top figure 1) where we expect the
composition of the core to have a larger effect. A planet that
has a larger iron content presumable mostly in its core, can accomodate
more of its mass in the central region and therefore have a smaller
radius compared to an iron-poor planet.
This type of planets (Super-Mercuries,  \citet{Valencia_et_al:2006})
exhibit comparatively large bulk densities as Mercury does in our
Solar System. If GJ876d had a core of 80\% by mass its radius would
only be 9200 km (or 1.45 $R_{\oplus}$) for pure Fe core and 9600 km (or 1.5
$R_{\oplus}$) for a sulfur enriched core. \\

Next consider the possibility of this planet having a water/ice layer
amounting to 20\% and 40\% of the total mass on top of a terrestrial core
(bottom figure 2).
Despite
ice and water being highly compressible (bulk modulus is 1-2 orders
of magnitude smaller respectively than for silicates), their densities at initial compression are low and the relation of the density gradient to density
is quadratic. Consequently, a large amount of water makes the planet
larger. If the Earth
had a substantial amount of water the radius would extend to 7100 km
(1.11$R_{\oplus}$) for 20\%
water content and 7600 km ($1.19R_{\oplus}$) for 40\% water content. 
Figure 2 shows the density structure and figure 3  - the
pressure-temperature structure 
for the different composition cases considered above.
The ocean planets have a layer of water above a layer of denser
ice due to the positive Clapeyron slope of ice VII (note the melting curve
of ice VII in figure 3). \citet{Hemley:1987} showed that ice transforms
gradually from ice VII to ice X with pressure and proposed an EOS for the
combined system. We have adopted those values for this calculation. Below the
ice layer the pressure is $\sim$ 150 GPa, beyond the transition pressure between
pv and ppv (P=125 GPa at T=2750K \citet{Tsuchiya:2004}), therefore the silicate mantle is made of
ppv+fmw --- perovskite and upper mantle silicate phases are absent. If GJ876d had a water/ice
layer that accounted for 20\% and a core that accounted for 33\% of
the total mass, the radius would be 11900 km (or 1.87 $R_{\oplus}$).
An additional 20\% of water would extend the radius 550 km more (to
1.95 $R_{\oplus}$).  These radii are 11 and 15\% larger respectively,
compared to the Earth-like composition radius result. Such planets will
be easier to detect when transiting their stars. Adding sulfur to the core
would extend the radius of both ocean planets by $\sim$130 km. The radius results for all compositions are
shown in Table 3.\\

In summary, the four bulk compositions produce differences in the
Super-Earth mean densities and radii that are large enough - from
3\% to 30\% in radius, to be observationally measurable. For example, if GJ876d
was transiting its star, MOST satellite photometry \citep{Rowe_et_al:2006} could achieve
3\% precision in its radius, with uncertainty dominated by the stellar
radius ($R=0.32R_{\odot}$). 

Our main result is that with a measurement of both mass and radius (even
with a precision of 10\%) we could certainly tell between an iron rich differentiated terrestrial planet
(R $<$ 9600 km) and a water rich (IMF$>$20\%) ocean planet (R $>$ 12000
km). This is because there is a limited family of compositions that would fit both. To illustrate
this point we have calculated the trade-off curve between the amount of ice (IMF)
and core (CMF) that would satisfy M=7.5$M_{\oplus}$ and R=11900km (see Figure
3).  The maximum of amount ices this planet could harbour is $\sim 49$\% with
CMF=51\% and no mantle . The
minimum could be zero if the planet had no core. This extreme case (no core)
is so unlikely that a radius detection of 11900km or more would necessarily mean that
the planet has a water layer. Hence, knowledge on just the radius and mass of a planet
-- in combination with internal structure models ---
could yield valuable structural and compositional information on the planet
and its formation environment.
\\

\section{Tidal heating effects}
The Super-Earth's
extreme close proximity to its star GJ876 means that even small
perturbations on its orbit might induce significant additional heating
from tides and change its internal structure. We check this by a simple
analysis, with a surface heat flux consisting of energy from radioactive
heat sources, secular cooling and tidal heating. We add different amounts of tidal
heating $Q_{tidal}$ to the heat flow $Q$ (see section 2.3) that translate in higher Rayleigh
numbers and thinner lithosperes and calculate the thermal
structure:
\begin{equation}
Q_{total}=Q + Q_{tidal} 
\end{equation}
 For the calculation of the effects of internal heating we assume
a temperature dependent viscosity.
  A simple analysis states that the energy dissipation
rate produced by an eccentric orbit in a synchronous planet is \citep{Murray_Dermott:satellite_tides}

\[
\frac{dE}{dt}=\frac{63}{4}\frac{e^{2}n}{\tilde{\mu}Q_d}\left(\frac{R_{p}}{a}\right)^{5}\frac{GM_{\star}^{2}}{a}\]
\\
where $G$ is the gravitational constant, $M_{\star}$ is the mass
of the star, $a$ is the semimajor axis, $n$ is the mean motion,
$\tilde{\mu}$ is the ratio of elastic to gravitational forces, $\tilde{\mu}\approx(10^{4}$km$/R_{p})^{2}$,
and $Q_d$ is the specific dissipation parameter. In other words the
energy available for tidal heating of the 7.5 $M_{\oplus}$ Earth-composition
planet GJ876d is
\begin{equation}
1.2\times10^{22}e^{2}~[W]
\end{equation}
where $Q_d=280$ as for the Earth \citep{Ray_et_al:2001}.
Figure 5 shows the effects of different values of $dE/dt$ on the
PT regime for this planet. In addition, figure 5
also shows the solidus for MgO, the lower mantle mineral with the
lowest melting temperature. By increasing the amount of tidal heating
the internal temperature rises and in some cases the lowermost
mantle might be partially molten. Our calculations show that large
increments in tidal heating cause small increments in internal temperature.
Thus, in order to melt the lowermost mantle, the tidal heating needs
to exceed $\sim6.8\times10^{17}$W for this 7.5 $M_{\oplus}$ planet (or
$\approx10^{17}$W/$M_{\oplus}$), meaning 2500 times the heat flow without
tidal heating. Even though the internal temperature is much higher than
without tidal heating, the radius only increases by $\sim$100 km. For planet GJ876d melting could
commence for eccentricites above 0.008. While we do not yet know the dynamic environment of this planet
very well, such eccentricities are not excluded \citep{Rivera_et_al:2005}.
We note that the newly discovered ${\sim}10~M_{\oplus}$ Super-Earth
HD69830b \citep{Lovis_et_al:2006} orbiting at 0.08~AU from a 0.86$M_{\odot}$ star is
also in a 3-planet system and appears to have a non-zero $e=0.10\pm 0.04$.
Hence its tidal heating could exceed $10^{17}~W/M_{\oplus}$. HD69830b is
expected to be mainly rocky \citep{Lovis_et_al:2006} and we estimate its radius at
1.84~$R_{\oplus}$ \citep{Valencia_et_al:2006}.
\\
\section{Uncertainties}
The uncertainties in the model come from uncertanties in the EOS and the
composition of the planet that needs to be known at high pressures and
temperatures.  
Any phase change has to be known a priori in
order to be incorporated in the model.  Therefore when we extrapolate to high pressures it
is possible that the material might change into a different phase that we can not
account for without experimental evidence.  If there are high-pressure phases
unaccounted for in the model, the radius here obtained for different compositions would be
an overestimate owing to the denser character of high-pressure
minerals. \citet{Benoit:1996} describes a transition of H$_2$O to ice XI at 300
GPa.  GJ876d with IMF=40\% reaches a maximum pressure of 350 GPa at the bottom
of the ice shell.  Therefore, the effects of ice XI would be to decrease the radius by a
few tens of kilometers for this planet and more for planets with larger masses
and IMF.\\

The likely compositions chosen for GJ876d yield
a central pressure between 2500 and 5000 GPa. Values that exceed
any laboratory experiment up to date \citep{Cohen_Gulseren_Hemley:2000}. The extrapolation
of EOS --- Vinet and 3rd order Birch-Murhnaghan --- is yet to be tested at these pressures. Nevertheless, the uncertainty in composition is
greater than errors in this extrapolation. The difference in radius from using
the two different EOS is $\sim 100$ km, less than the difference 
due to the different compositions considered here.
There are also uncertainties in the determination of the parameters in the EOS
that translate into uncertainties in the internal structure. We considered two
different data sets for the EOS of pure Fe \citep{Uchida_et_al:2001,Mao:1990}
and find that the difference in radius is only of $\sim$65 km for the planet
dominanted by core composition. 
\\
Finally, the different treatments in viscosity (isoviscous vs. temperature
dependent) only yield a difference in radius of
$\sim$20 km, though the temperature regime is different for the four
compositions described herein. Our lack of knowledge on the temperature
structure is offset by the small temperature effects in the total radius of the
planet. This means that despite the uncertainties in the thermal model (viscosity
assumptions and heat flow scaling assumptions) the radius is a robust
parameter. Therefore we place an estimate on the error in radius of $\sim$200
km due to the lack of knowledge associated with the EOS and in the
thermal profile.

\section{Conclusions and Discussion}
In conclusion, the new surveys and improved detection techniques have
opened up the study of a new class of objects, Super-Earths, that brings
us closer to characterizing the exoplanets most similar to Earth.
In particular, we can model the internal structure of the first discovered
Super-Earth by looking at likely compositions. We find that the
radius varies between $\sim$9200 and 12500 km. If the orbital geometry
allowed a transit follow up to GJ876d, the expected flux drop would
be of $1.9-3.6\times10^{-3}$, large enough to be observed.
If observed with a large-aperture telescope \citep{Holman_et_al:2005} or from space,
e.g. with MOST \citep{Rowe_et_al:2006}, a transiting planet like GJ876d would
yield a radius determination with $\sim3$\% precision and one would be able to distinguish between all four scenarios/models presented
by us here, especially between an iron rich differentiated and an ocean
planet. Moreover, for a given mass there is a radius that delimits the boundary
between ocean planets and terrestrial planets. If GJ876d radius was larger
than 12000km, it would indicate it is an ocean planet.

 With the upcoming space mission \emph{Kepler} that ability
will improve, especially with expected advances in RV measurements
(hence planet masses) for \emph{Kepler's} fainter targets. Therefore,
when radii measurements of terrestrial planets become available, we
can put constraints on the bulk composition of a planet (given
its mass) and begin to understand the conditions on distant terrestrial
planets. A following article will treat this problem by showing in
a simple manner the relation between radius, mass and composition.\\ %

\section{Acknowledgements}
We would like to thank the anonymous reviewer for his/her useful comments.
This work was supported under grant NSF grant EAR-0440017.  Diana Valencia acknowledges support from the Harvard Origins
of Life Initiative.


\begin{deluxetable}{ccccccccc} 
\rotate
\tablewidth{0pt}
\tablecolumns{9}
\tablecaption{Compositional data}
\tablehead{
\colhead{layer} & \colhead{composition} & \colhead{$\rho_0$} (kg/m$^3$) &
\colhead{$K_0$} (GPa) & \colhead{$K_0^\prime$} & \colhead{$\gamma_0$} &
\colhead{$q$} & \colhead{$\theta_0$} & \colhead{ref}}

\startdata

\it ocean  & H$_2$O &  998.23 &   2.18  &  -  &  -  &
- &  - & 1 \\
\it shell & ice VII+X &  1463  &   2.308  &  4.532 &   1.2  &
1. &  1470$^3$ & 2,3 \\ 
\multicolumn{5}{c}\nodata & a$_1$=-4.2$\times 10^-4$ & a$_2$=1.56 $\times 10^-6$  & b = 1.1 & 2  \\ \tableline	
\it upper & ol &  3347.  &   126.8  &  4.274 &   0.99 &	2.1 &  809 & 4	\\ 
\it mantle & wd+rw &  3644.  &   174.5  &  4.274 &   1.20 &	2.0 &  908b & 4 \\ 	\tableline
\it lower & pv+fmw &  4152.  &   223.6  &  4.274 &   1.48 &	1.4 &  1070 & 4	\\ 
\it mantle  & ppv+fmw & 4270.  &   233.6  &  4.524 &   1.68 &	2.2 &
1100 & 5 \\     \tableline
\it core & Fe & 8300  &   160.2  &  5.82  &   1.36 &	0.91 & 998 & 6,7\\
 & Fe$_{0.8}$(FeS)$_{0.2}$ & 7171 & 150.2 &  5.675 &  1.36 & 0.91 & 998 & 6,7\\
\enddata
\tablerefs{Compositional data used in each shell of the model has been taken from
different sources and refit to the Vinet EOS:
(1) \citet{Stewart_Ahrens:2005} --- Rankine-Hugoniot EOS 
(2) \citet{Hemley:1987}
(3) \citet{Fei_et_al:1993}
(4) \citet{Stixrude_Lithgow:2005} --- A reuss average was performed according
to a mixture of 10\% Fe and 30\% fmw for the bulk modulus. The thermodynamic
parameters ($\gamma0$, $q$ and $\theta_0$) were taken from the most dominant
phase in each shell.
(5) \citet{Tsuchiya:2004} -- an increase in density was needed to account for
20\% Fe in ppv according to $\partial ln / \partial x=0.3 $ where $x$ is the iron
content \citep{Mao:2004}. A Reuss average was then used with fmw.
(6) \citet{Williams_Knittle:1997} --- we also tried \citet{Uchida_et_al:2001}
parameters for density and bulk modulus and found a discrepancy in the radius
of only 60 km for a compositon of CMF=80\%.
(7) \citet{Uchida_et_al:2001}}
\end{deluxetable}

\begin{deluxetable}{ccc}
\tablewidth{0pt}
\tablecolumns{3}
\tablecaption{Phase boundaries in silicate mantle}
\tablehead{\colhead{transition} & \colhead{boundary} & \colhead{} }
\startdata
ol $\rightarrow$ wd+rw \phm{$\rightarrow$} & T=400P-4287 &  {} \\ \tableline 
rw $\rightarrow$ pv+fmw\phm{$\rightarrow$} & P=22.6  & if T $>$ 1750K \\
{} & T=13573-500P & if T $\le$ 1750K \\ \tableline
pv+fmw $\rightarrow$ ppv+fmw \phm{$\rightarrow$}  & T=133P-1392 & {}\\ 
\enddata
\tablecomments{Table showing the different phase boundaries between the silicate
mantle minerals used in this study. P is in GPa and T is in K.}
\end{deluxetable}

\bigskip

\begin{deluxetable}{crr}
\tablewidth{0pt}
\tablecolumns{3}
\tablecaption{GJ875d radius}
\tablehead{\colhead{RADIUS}   & \colhead{Fe} & \colhead{Fe$_{0.8}$(FeS)$_{0.2}$}}  
\startdata
Earth-like & 10786 km  & 10914 km \\
Rocky CMF=80\% & 9228 km  & 9580 km\\ 
Ocean IMF=20\% & 11890 km & 12014 km\\ 
Ocean IMF=40\% & 12448 km & 12576 km\\
\enddata
\tablecomments{Radius results for the different four bulk compositions for
a pure Fe core and a core with Fe and S. The only appreciable difference of
400 km happens in a core-dominated planet.}
\end{deluxetable}


\begin{figure}
\figurenum{1}\begin{center}
\includegraphics[scale=0.50]{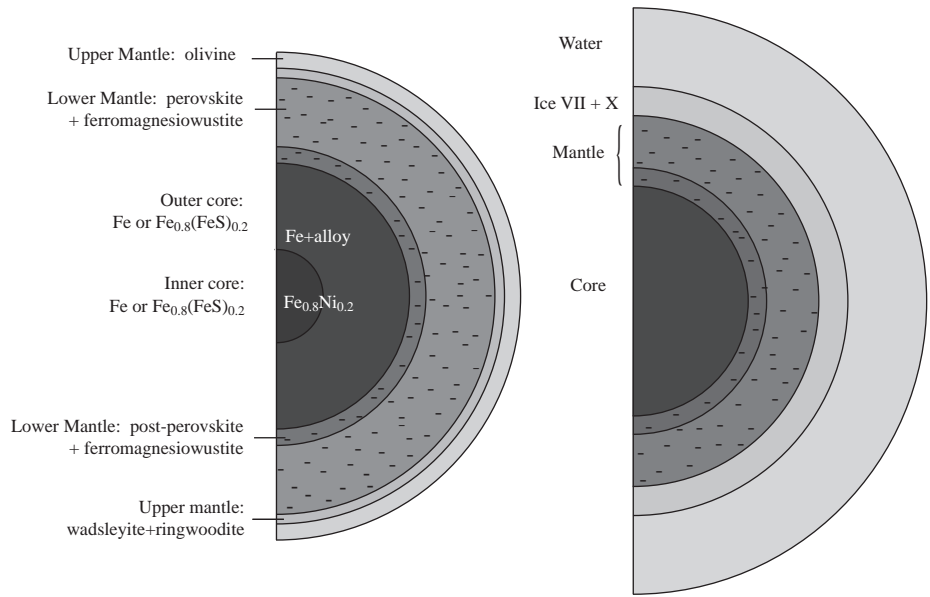}\end{center}
\caption{\footnotesize Schematic representation of the model. To calculate
 the internal structure of Super-Earths we assume a similar composition to the Earth's: A dense core of pure Fe or Fe$_{0.8}$(FeS)$_{0.2}$
 as end member cases (the Earth has an outer core of Fe plus some unknown
 alloy and the solid inner core has Fe and Ni); a lower mantle composed of
 two silicate shells (ppv+fmw, pv+fmw); an upper mantle composed of two
 silicate shells (wd+rw, ol).  The thickness of the shells will depend on the
 PT profile for the planet and the amount of mass in the core.  An ocean
 planet - right - will have an additional water/ice layer above the rocky
 core. }
\end{figure}

\begin{figure}
\figurenum{2}
\begin{center}\includegraphics[%
  width=0.75\linewidth]{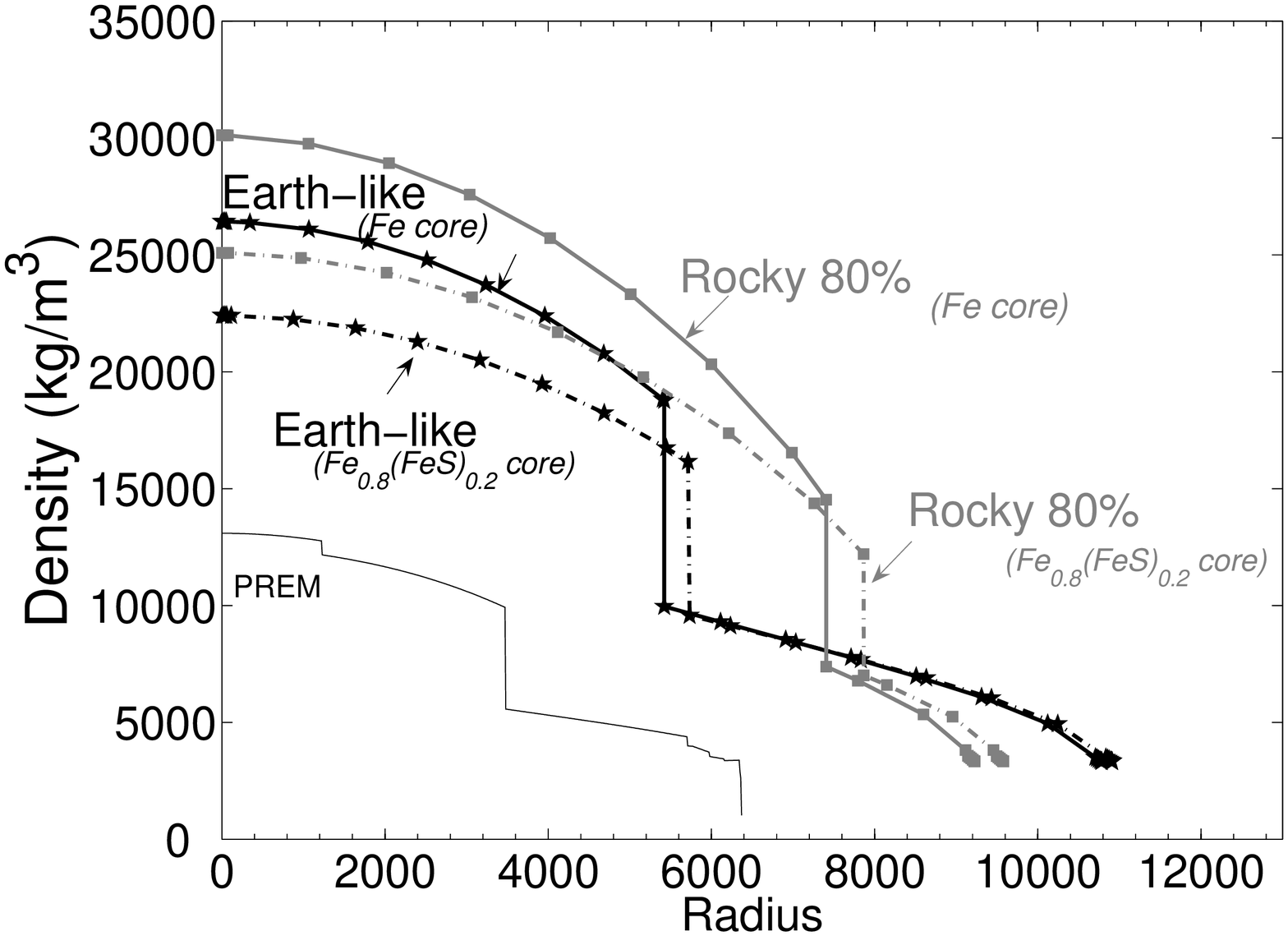}\end{center}
\begin{center}\includegraphics[%
  width=0.75\linewidth]{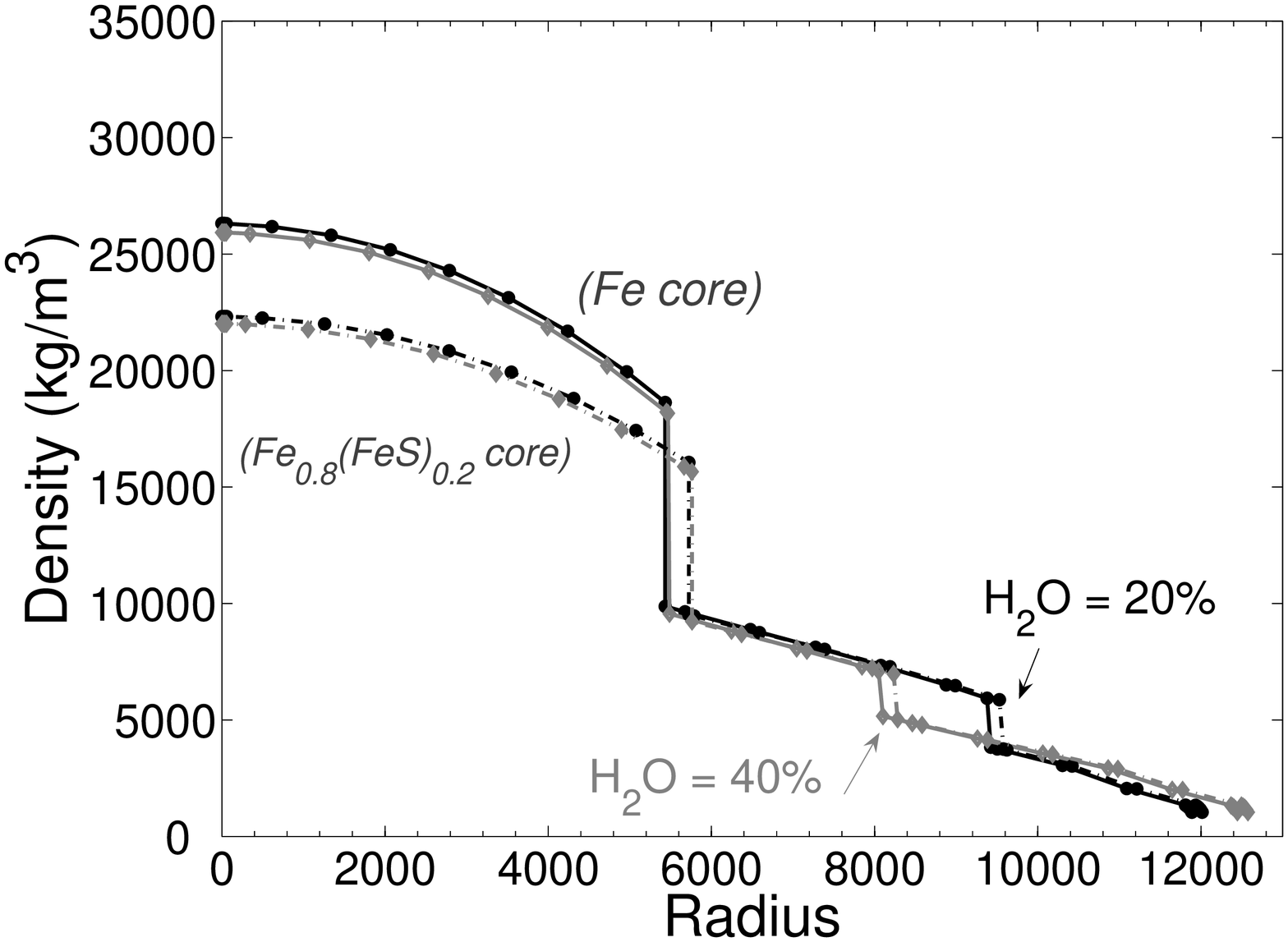}\end{center}
\caption{\footnotesize Internal Structure
of exoplanet Gliese 876d: Density Profile.
Four different compositions are illustrated. The surface
is on the right and the centre of the planet is on the left. The solid lines
show the cases for which the composition of the core is taken to be pure
Fe. Dashed lines are for the case of Fe$_{0.2}$(FeS)$_{0.8}$. Lines with stars
shows the internal structure of GJ876d if the composition was Earth-like.
Square symbols show the density profile if this planet had
80\% of the mass in the core. Circles show the
structure if this planet had formed outside the snow line and retained
20\% of its mass as a water/ice layer. Diamonds shows the density structure if
GJ876d had retained 40\% of water/ice.  A Preliminary Reference model for
Earth (PREM -- \citet{PREM} ) is shown
for reference. }
\end{figure}

\begin{figure}
\figurenum{3}
\begin{center}\includegraphics[%
 width=0.95\linewidth]{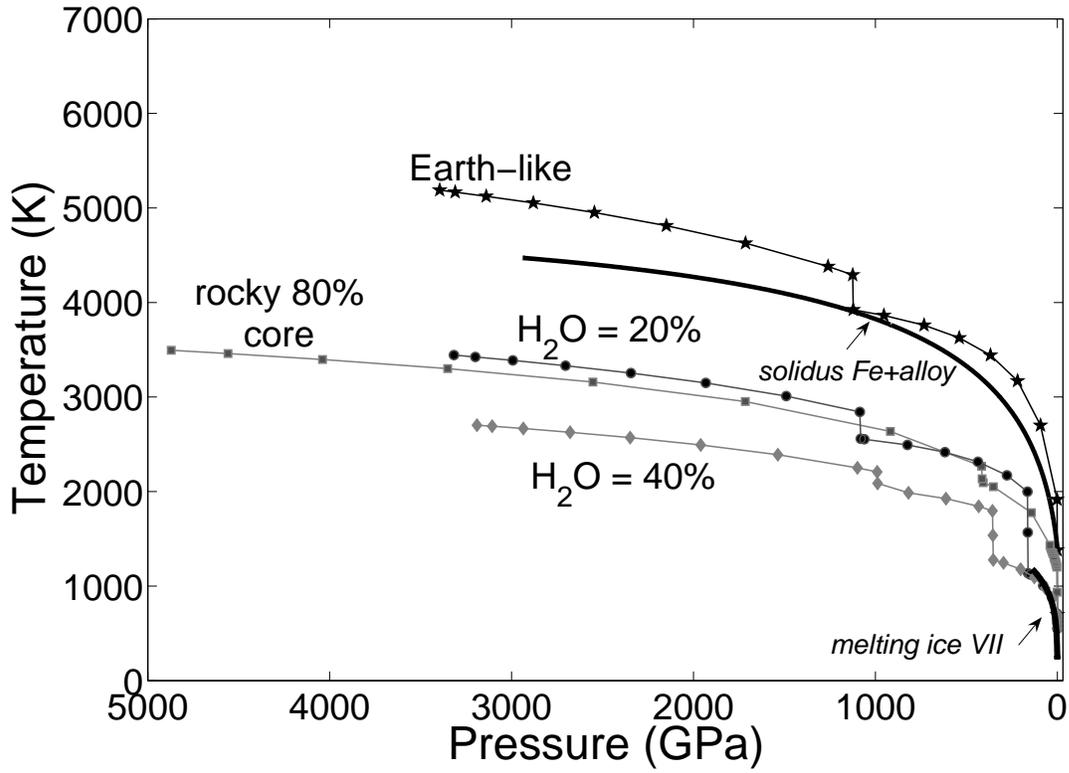}\end{center}
\caption{\footnotesize 
Internal Structure of exoplanet Gliese 876d:
Pressure-Temperature Structure.  The PT structure of
GJ876d is illustrated considering the four compositions presented
in this study and pure Fe in the core. The ocean-compositions cross the ice VII solidus at
$\sim$55 GPa which leaves means the water layer is 1200 km deep for both ocean
planets (IMF=20\% and IMF=40\%).}
\end{figure}

\begin{figure}[h]
\figurenum{4}
\begin{center}\includegraphics[%
   width=0.95\linewidth]{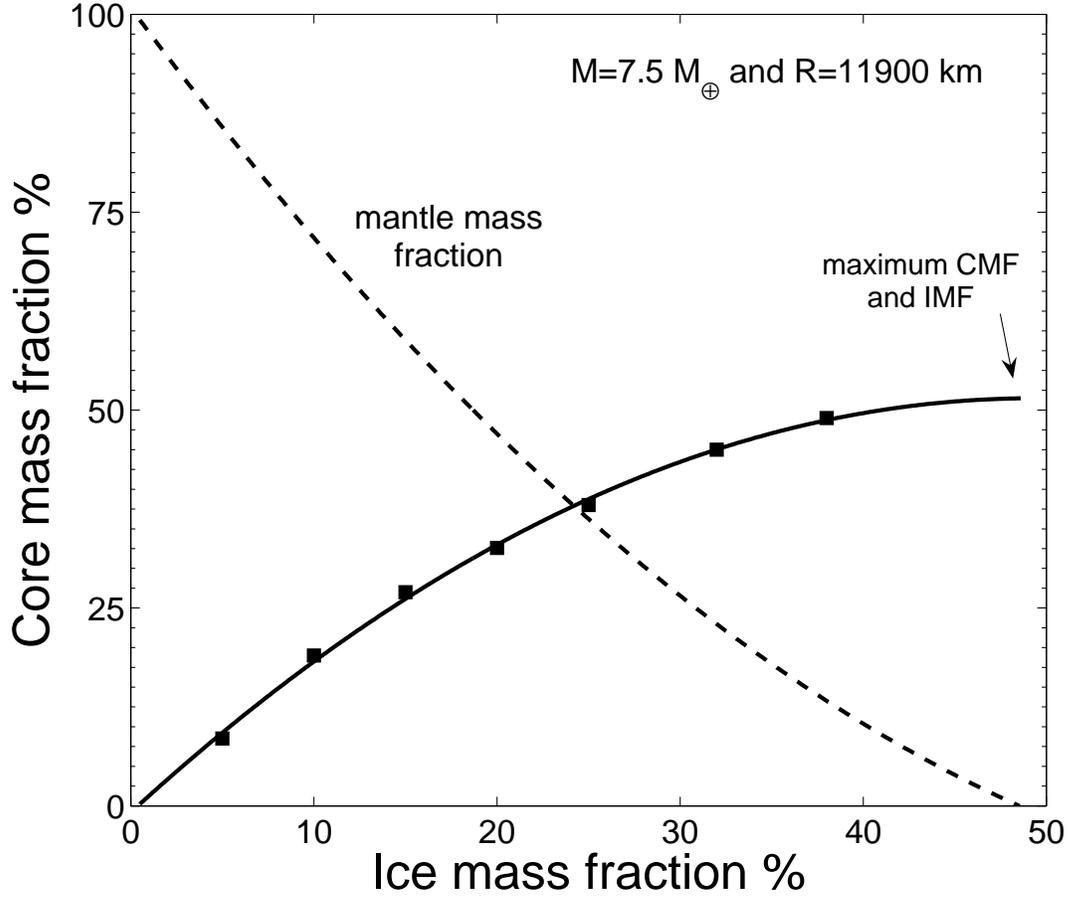}\end{center} 
\caption{\footnotesize Trade-off curve beetween CMF and IMF. For a planet with $M=7.5 M_{\oplus}$ and $R=11900$ km the trade-off
 curve between the amount of ices and amount of mass in the core is depicted
 as a solid line. The squares correspond to the different data points.  The
 dashed line shows the amount of mass in the mantle. The maximum amount of
 ices this planet could harbour is 48.6\% with CMF=51.4\% and no mantle. A
 planet this size can only be rocky if it had no core (IMF=0 when CMF=0). }
\end{figure}

\begin{figure}[h]
\figurenum{5}
\begin{center}
\includegraphics[
  width=0.95\linewidth]{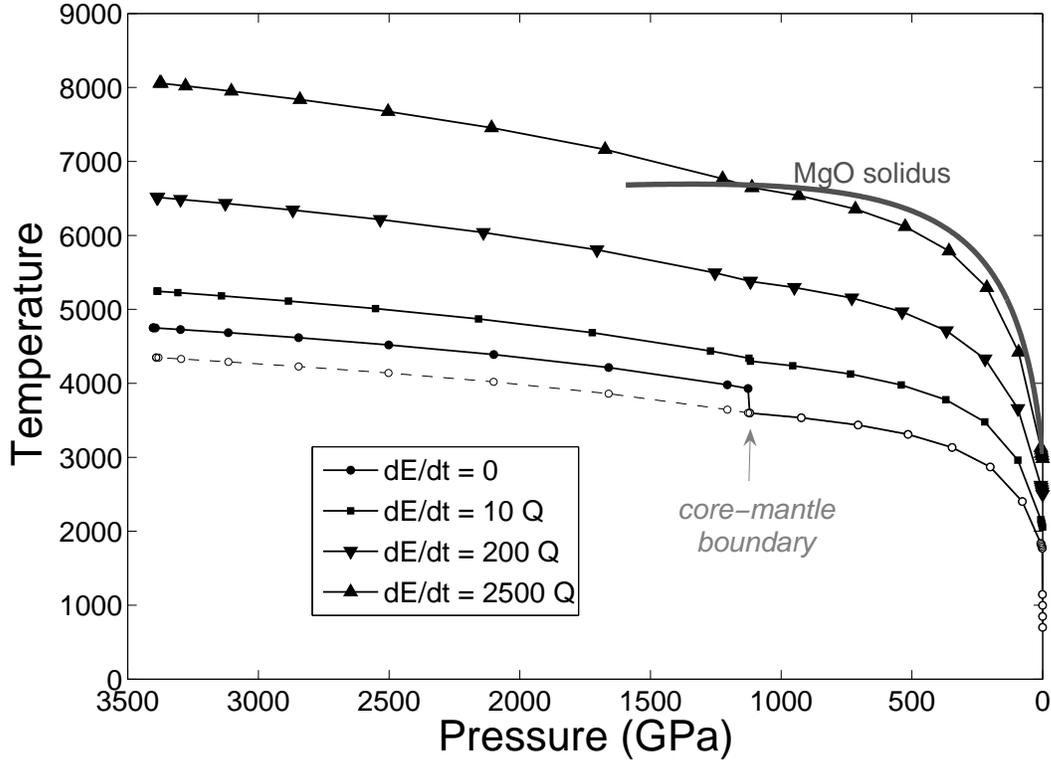}\end{center}
\caption{\footnotesize Effects of tidal heating on PT structure. PT structure for a 7.5 $M_{\oplus}$ planet experiencing different
degrees of tidal heating.  Dots: no tidal heating. Squares:
tidal heating that is 10 times the heat flow withouth tidal heating ($Q$) or
total heat flow $Q_{total}$=3000 TW. Downward-pointing
triangles: 200 times or  $Q_{total}$=5500 TW. Upward- pointing triangles:
2500 times or  $Q_{total}$=68000 TW. The solidus for MgO is crossed in the mantle when
the tidal heating is 2500 times the radioactive and secular cooling contributions. Dotted
lines refer to a thermal model without a lower boundary layer in the
mantle. As the tidal heating increases the lower boundary layer becomes
negligible.}
\end{figure}


\bibliographystyle{apj}
\bibliography{/home/valencia/Documents/MyPapers/proj}

\end{document}